\documentclass[a4paper,11pt]{article}
\usepackage{pos}

    \usepackage{array}
    \usepackage{multirow}

\newcommand{\PreserveBackslash}[1]{\let\temp=\\#1\let\\=\temp}
\newcolumntype{C}[1]{>{\PreserveBackslash\centering}p{#1}}
\newcolumntype{R}[1]{>{\PreserveBackslash\raggedleft}p{#1}}
\newcolumntype{L}[1]{>{\PreserveBackslash\raggedright}p{#1}}

\title{Tetraquark Interpretation and Production Mechanism of the Belle $Y_b (10750)$-Resonance}

\author[a]{Ahmed Ali}
\author[b,c]{Luciano Maiani}
\author*[d]{Alexander Parkhomenko}
\author[b,e]{Wei Wang}

\affiliation[a]{Deutsches Elektronen-Synchrotron DESY, \\ 
  Street number, D-22607 Hamburg, Germany}

\affiliation[b]{T.\,D.~Lee Institute, Shanghai Jiao Tong University, \\ 
  Shanghai 200240, China} 

\affiliation[c]{Dipartimento di Fisica and INFN, Sapienza Universit\`{a} di Roma, \\  
  Piazzale Aldo Moro 2, I-00185 Roma, Italy} 

\affiliation[d]{Department of Theoretical Physics, P.\,G.~Demidov Yaroslavl State University, \\ 
  Sovietskaya 14, 150003 Yaroslavl, Russia}

\affiliation[e]{INPAC, SKLPPC, MOE KLPPC, School of Physics and Astronomy, Shanghai Jiao Tong University, \\  
  Shanghai 200240, China} 

\emailAdd{ahmed.ali@desy.de}
\emailAdd{luciano.maiani@cern.ch}
\emailAdd{parkh@uniyar.ac.ru}
\emailAdd{wei.wang@sjtu.edu.cn}

\renewcommand{\printHeadAuthors}{A.~Ali et al.}

\abstract{
Recently, the Belle Collaboration has updated the analysis of the cross sections for the processes
$e^+ e^- \to \Upsilon(nS)\, \pi^+ \pi^-$ ($n = 1,\, 2,\, 3$) in the $e^+ e^-$ center-of-mass energy
range from 10.52 to 11.02~GeV. A new structure, called $Y_b (10750)$, 
with the mass $M (Y_b) = (10752.7 \pm 5.9^{+0.7}_{-1.1})$~MeV and the Breit-Wigner width 
$\Gamma (Y_b) = (35.5^{+17.6 +3.9}_{-11.3 -3.3})$~MeV was observed.  
We interpret $Y_b (10750)$ as a compact $J^{PC} = 1^{--}$ state with a dominant tetraquark component.
The mass eigenstate $Y_b (10750)$ is treated as a linear combination of the diquark-antidiquark 
and $b \bar b$ components due to the mixing via gluonic exchanges shown recently to arise in the limit 
of large number of quark colors. The mixing angle between $Y_b$ and $\Upsilon(5S)$ can be estimated  
from the electronic width, recently determined to be $\Gamma_{ee} (Y_b) = (13.7 \pm 1.8)$~eV.  
The mixing provides a plausible mechanism for $Y_b (10750)$ production in high energy collisions 
from its $b \bar b$ component and we work out the Drell-Yan and prompt production 
cross sections for $p p \to Y_b (10750) \to \Upsilon (nS)\, \pi^+ \pi^-$ at the LHC.  
The resonant part of the dipion invariant mass spectrum in $Y_b (10750) \to \Upsilon (1S)\, \pi^+ \pi^-$ and 
the corresponding angular distribution of $\pi^+$-meson in the dipion rest frame are presented. 
}

\FullConference{%
  40th International Conference on High Energy physics - ICHEP2020\\
  July 28 - August 6, 2020\\
  Prague, Czech Republic (virtual meeting)
}

\begin{document}
\maketitle

\section{Introduction} 

In 2019, the Belle Collaboration has presented the updated analysis of the cross sections 
for $e^+ e^- \to \Upsilon(nS)\, \pi^+ \pi^-$ ($n = 1,\, 2,\, 3$) in the electron-positron 
center-of-mass energy range from 10.52 to 11.02~GeV~\cite{Abdesselam:2019gth}.
In addition to the already known structures $\Upsilon (10860)$ and $\Upsilon (11020)$, 
they reported the observation of a lower-mass resonance called $Y_b (10750)$ with the 
global significance of~$5.2\sigma$. 
The measured masses and decay widths (in~MeV), and ranges of $\Gamma_{ee} \times {\cal B}$ 
(in~eV) of all three structures 
are borrowed from~\cite{Abdesselam:2019gth} and presented in Table~\ref{tab:Belle-data-1-2019}. 

The combined analysis of the BaBar and Belle data on the ratio $R_{b \bar b}$ has been recently 
undertaken~\cite{Dong:2020tdw} with an account of the coherent sum of $\Upsilon (10860)$, 
$\Upsilon (11020)$, and $Y_b (10750)$ and a continuum amplitude, proportional to $1/\sqrt s$, 
where $\sqrt s$ is the center-of-mass energy of the $e^+ e^-$-pair. Masses, Breit-Wigner decay 
widths, leptonic partial decay widths, and relative phases are fit parameters of the $R_{b \bar b}$ 
lineshape. One gets a number of solutions for the partial electronic widths (mathematically 
8~solutions are expected), which differ in other parameters~\cite{Dong:2020tdw}. 
Most of these solutions are likely unphysical except the solution, in which the electronic width 
of~$Y_b$ is given as\footnote{
In the analysis~\cite{Ali:2019okl} on which this talk is based we used the preliminary results 
of X.-K. Dang et al.~\cite{Dong:2020tdw} presented by Chang-Zheng Yuan in XVth Rencontres du Vietnam  
(September 22nd -- 28th, 2019, Quy Nhon, Vietnam).}: 
\begin{eqnarray} 
\Gamma_{ee} (Y_b (10750)) &=& \left ( 13.7 \pm 1.8 \right ) {\rm eV}. 
\label{mixval}
\end{eqnarray}
The resulting masses and decay widths of the three resonances are found to be in agreement 
with the ones obtained from the $R_{\Upsilon\, \pi^+ \pi^-}$ scan~\cite{Abdesselam:2019gth}. 
However, $Y_b (10750)$ is open to a number of interpretations to be tested in the existing 
and future experiments. 

The conventional interpretation of $Y_b (10750)$ is that $Y_b (10750)$ and $\Upsilon (10680)$ 
are a mixture of $\Upsilon (5S)$ and $\Upsilon (4D)$~\cite{Badalian:2009bu}. In~\cite{Ali:2019okl}, 
we interpret $Y_b (10750)$ as a $J^{PC} = 1^{--}$ tetraquark candidate, whose dominant component~$Y_b^0$ 
consists of a colored diquark-antidiquark pair $[b q]_{\bar 3_c} [\bar b \bar q]_{3_c}$, bound 
in the $SU (3)$ antitriplet-triplet representation~\cite{Jaffe:2003sg,Maiani:2004vq}. However, 
it can have a small $b \bar b$ component due to the mixing via gluonic exchanges. On the other hand, 
$\Upsilon (10860)$ and $\Upsilon (11020)$, which are dominantly radial $b \bar b$ excitations, 
$\Upsilon (5S)$ and $\Upsilon (6S)$, respectively, also have a small diquark-antidiquark 
component~$Y_b^0$ in their content. Due to the proximity of the mass eigenstates $Y_b (10750)$ 
and $\Upsilon (10860)$, we consider that the mixing is dominantly between~$Y_b^0$ and $\Upsilon (5S)$. 
This also provides a plausible interpretation of some anomalous features measured 
in the $\Upsilon (10860)$-decays.

Mixing between a bottomonium and hidden-beauty tetraquark, anticipated in~\cite{Knecht:2013yqa}, 
was shown in~\cite{Maiani:2018pef}, to be induced at the level of non-planar diagrams in the 
large-$N_c$ limit.  
Albeit suppressed at large~$N_c$ by the exponential factor $e^{-N_c/2}$, when extrapolated 
back to $N_c = 3$ one finds a result not dissimilar from $f \sim N_c^{-3/2}$. Thus, a production 
in the $e^+ e^-$-annihilation of resonances such as $Y_b (10750)$, in addition to the bottomonium 
spectral lines and with a small $\Gamma_{ee}$, is a significant signature of tetraquarks.  

\begin{table}[tb]
\caption{
Measured masses and decay widths (in MeV), and ranges of $\Gamma_{ee} \times {\cal B}$ (in eV) 
of the $\Upsilon (10860)$, $\Upsilon (11020)$, and the new structure $Y_b (10750)$. 
The first uncertainty is statistical and the second is systematic.  
}
\label{tab:Belle-data-1-2019} 
\begin{center}
\hspace{-4mm}
\begin{tabular}{cccc} 
\hline
State        & $\Upsilon (10860)$    &     $\Upsilon (11020)$     &     $Y_b(10750)$     \\ \hline      
Mass  & {\footnotesize $10885.3 \pm 1.5^{+2.2}_{-0.9}$} & {\footnotesize $11000.0^{+4.0  +1.0}_{-4.5 -1.3}$} & {\footnotesize  $10752.7 \pm 5.9^{+0.7}_{-1.1}$} \\
Width & {\footnotesize  $36.6^{+4.5 +0.5}_{-3.9 -1.1}$} & {\footnotesize     $23.8^{+8.0 +0.7}_{-6.8 -1.8}$} & {\footnotesize $35.5^{+17.6 +3.9}_{-11.3 -3.3}$} \\
\hline  
$\Upsilon(1S) \pi^+ \pi^-$ & {\footnotesize $0.75 - 1.43$} & {\footnotesize $0.38 - 0.54$} & {\footnotesize $0.12 - 0.47$} \\[1mm] 
$\Upsilon(2S) \pi^+ \pi^-$ & {\footnotesize $1.35 - 3.80$} & {\footnotesize $0.13 - 1.16$} & {\footnotesize $0.53 - 1.22$} \\[1mm] 
$\Upsilon(3S) \pi^+ \pi^-$ & {\footnotesize $0.43 - 1.03$} & {\footnotesize $0.17 - 0.49$} & {\footnotesize $0.21 - 0.26$} \\[1mm] \hline 
\end{tabular}
\end{center} 
\end{table}


\section{Bottomonium-Tetraquark Mixing Formalism}

Let us define the tetraquark states $Y_b^I$ ($I = 0,\, 1$) in the isospin basis as    
$Y_b^0 \equiv \left ( Y_{[bu]} + Y_{[bd]} \right )/\sqrt 2$ and
$Y_b^1 \equiv \left ( Y_{[bu]} - Y_{[bd]} \right )/\sqrt 2$. 
Their mass difference due to the isospin breaking can be ignored. Their production is possible 
via the isosinglet $b \bar b$-component, so only the $Y_b^0$-state is considered.  

Experimentally observed mass differences are $M [\Upsilon (10860)] - M [Y_b (10750)] \simeq 133$~MeV 
and $M [\Upsilon (11020)] - M [Y_b (10750)] \simeq 247$~MeV, so the mixing between $\Upsilon (10860)$ 
and $Y_b (10750)$ states should be more pronounced:     
\begin{equation}
\left (
\begin{array}{c} 
      Y_b (10750) \\
 \Upsilon (10860)
\end{array}
\right ) 
=
\left ( 
\begin{array}{rr} 
  \cos\tilde\theta & \sin\tilde\theta \\
- \sin\tilde\theta & \cos\tilde\theta
\end{array}
\right ) 
\left (
\begin{array}{c} 
         Y_b^0 \\
 \Upsilon (5S)
\end{array}
\right ) . 
\label{eq:state-rotation}
\end{equation}
The mixing angle~$\tilde\theta$ is estimated phenomenologically. 
In general, the mixing can be easily generalized to the case of all three states. 
This mixing relates $\Gamma_{ee} [Y_b (10750)]$ and $\Gamma_{ee} [\Upsilon (5S)]$:  
\begin{equation}
\frac{\Gamma_{ee} [Y_b (10750)]}{\Gamma_{ee} [\Upsilon (10860)]} = 
\tan^2\tilde\theta \left [ \frac{M [\Upsilon (10860)]}{M [Y_b (10750)]} \right ]^4 
\simeq 1.04\, \tan^2\tilde\theta .   
\label{eq:Gamma-ee-Yb-5S}
\end{equation}
LHS of this equation can be determined numerically, using  
$\Gamma_{ee} [\Upsilon (10860)] = (310 \pm 70)$~eV~\cite{Zyla:2020zbs}   
and $\Gamma_{ee} [Y_b (10750)] = (13.7 \pm 1.8)$~eV.  
The estimate of the mixing angle is  
$\tan^2\tilde\theta = 0.044 \pm 0.006$ and $\tilde\theta \simeq 12^\circ$~\cite{Ali:2019okl}. 
This supports the prediction from the large-$N_c$ approach that the mixing angle between 
a pure bottomonium and hidden-bottom tetraquark state is relatively large.

\section{Production Cross Sections at the LHC} 

Hadroproduction cross sections for $\Upsilon (5S)$ and $\Upsilon (6S)$ in $p \bar p$ (Tevatron) 
and $p p$ (LHC) collisions were calculated using the NRQCD framework~\cite{Ali:2013xba}, 
assuming  a factorization ansatz to separate the short- and long-distance effects.  
Cross-sections for $Y_b (10750)$ are scaled from the ones for $\Upsilon (5S)$, since in both cases 
the production takes place via the $b \bar b$-component of $Y_b (10750)$. 
The cross-section ratio is determined by the mixing angle~\cite{Ali:2019okl}:  
\begin{eqnarray}
&& \hspace{-11mm}  
\frac{\sigma ( p p \to Y_b (10750) + X)\, {\cal B}_f (Y_b (10750))}
     {\sigma ( p p \to \Upsilon (10860) + X)\, {\cal B}_f (\Upsilon (10860))}
\nonumber \\
&& \hspace{-11mm} \simeq 
\frac{\Gamma_{ee} (Y_b (10750))\, {\cal B}_f (Y_b (10750))} 
     {\Gamma_{ee} (\Upsilon (10860))\,  {\cal B}_f (\Upsilon (10860))} 
\simeq 1.04\, \tan^2\tilde\theta\, 
\frac{{\cal B}_f (Y_b (10750))}{{\cal B}_f (\Upsilon (10860))} , 
\label{eq:Yb-Upsilon}
\end{eqnarray} 
where ${\cal B}_f (Y_b (10750)) = {\cal B} (Y_b (10750) \to \Upsilon (nS) \pi^+ \pi^-)$ and 
${\cal B}_f (\Upsilon (10860)) = {\cal B} (\Upsilon (10860) \to \Upsilon (nS) \pi^+ \pi^-)$ 
with $n = 1,\, 2,\, 3$.   
RHS of this equation has been measured by Belle~\cite{Abdesselam:2019gth}.  
Absolute cross sections for $\Upsilon (10860)$ are estimated in NRQCD~\cite{Ali:2013xba}:  
%
\begin{equation}
\sigma (p p \to \Upsilon (10860) + X) = \sum _Q \sigma_Q 
= \sum_Q \int dx_1 dx_2 \sum_{i,j} f_i (x_1)\, f_j (x_2)\,    
\hat\sigma \left ( i j \to \langle \bar b b \rangle_Q + X \right ) \langle O [Q] \rangle ,  
\label{eq:sigma-pp}
\end{equation} 
where $f_i (x_1)$ is the parton distribution function (PDF) of a generic $i$-th parton inside a proton,     
the label $Q = {}^{2S+1} L_J^c$ denotes the $b \bar b$-pair quantum number (color~$c$, spin~$S$, and 
orbital~$L$ and total~$J$ angular momenta), $\langle O [Q] \rangle$ are long-distance matrix elements 
(LDMEs), and $\hat\sigma = \sigma/\langle O [Q] \rangle$ is a normalized partonic cross section. 
Leading-order partonic processes for the $S$-wave configurations are $g + g \to \Upsilon [^3 S_1^1] + g$,
$g + g \to \Upsilon [^1 S_0^8,\; ^3 S_1^8] + g$, $g + q \to \Upsilon [^1 S_0^8,\; ^3 S_1^8] + q$, 
and $q + \bar q \to \Upsilon [^1 S_0^8,\; ^3 S_1^8] + g$. 
Details of the calculations are presented in~\cite{Ali:2019okl}.   
 
Numerical estimates for hadroproduction and Drell-Yan cross sections at the LHC are collected 
in Table~\ref{tab:fullresult}. Total cross sections (in~pb) at $\sqrt s = 14$~TeV are calculated 
for $p p \to Y_b (10750) + X \to \Upsilon (nS) (\to \mu^+\mu^-)\, \pi^+\pi^- + X$ $(n = 1,\, 2,\, 3)$ 
at the LHC, assuming the transverse momentum range $3~{\rm GeV} < p_T < 50~{\rm GeV}$.    
Rapidity ranges are $|y| < 2.5$ for ATLAS and CMS (called LHC~14) and $2.0 < y < 4.5$ for the LHCb.   
Error estimates in the QCD production are from the variation of the central values 
of the Color-Octet LDMEs and the various decay branching ratios, as discussed in~\cite{Ali:2013xba}.  
Contributions from $\Upsilon (1S,\, 2S,\, 3S)$ are added together 
in the Drell-Yan production mechanism~\cite{Ali:2011qi}. 
To estimate the expected number of events, we use 1~pb for the cross section, which lies in the middle 
of the indicated ranges, yielding $O (10^4)$ signal events at the LHCb, and an order of magnitude larger 
for the other two experiments, ATLAS and CMS. In addition to the production due to the mixing mechanism, 
there may be direct production of tetraquarks, which would add incoherently. Thus, the numbers 
in Table~\ref{tab:fullresult} give lower bounds to the expected $Y_b (10750)$ production in proton collisions.

\begin{table*}
\caption{
Total cross sections (in~pb) for the processes
$p p \to Y_b (10750) \to (\Upsilon(nS) \to \mu^+\mu^-)\, \pi^+\pi^-$ $(n = 1,\, 2,\, 3)$  
at the LHC ($\sqrt s = 14$~TeV), assuming the transverse momentum range $3~{\rm GeV} < p_T < 50~{\rm GeV}$.  
}
\label{tab:fullresult}
\begin{center}
\begin{tabular}{lccc|cccc}
\hline\hline
 &
 &
 QCD (gg)
 &
 & 
 Drell-Yan
 &
 \\
 & $n = 1$ & $n = 2$ & $n = 3$
 & DY\\\hline 
LHC 14
 & [ 0.29,  3.85]
 & [ 0.70,  4.78]
 & [ 0.45,  3.10] 
 & [0.002, 0.004]
\\
LHCb 14
 & [ 0.08,  1.21]
 & [ 0.20,  1.51]
 & [ 0.13,  0.99] 
 & [0.001, 0.002]
\\
 \hline\hline
\end{tabular} 
\end{center}
\end{table*}

\section{Dipion Invariant Mass and Angular Distributions} 

The total amplitude ${\cal M} = \sum_I {\cal M}_I$ of the $Y_b (10750) \to \Upsilon (nS)\, P P^\prime$ 
decay (here, $P^{(\prime)}$ denotes a pseudoscalar meson), and is the sum over possible isospin states 
${\cal M}_I = {\cal S},\ {\cal D},\ {\cal D}',\ {\cal D}''$~\cite{Ali:2010pq}.   
Here, ${\cal S}$ is the $S$-wave amplitude for $P P^\prime$ system and 
${\cal D}$, ${\cal D}'$ and ${\cal D}''$ are the $D$-wave amplitudes.    
For the $\pi^+ \pi^-$-pair in the $Y_b (10750) \to \Upsilon (1S)\, \pi^+\pi^-$ decay, 
the isospin is $I = 0$ and the scalar $\sigma = f_0 (500)$- and $f_0 (980)$- and the tensor 
$f_2 (1270)$-resonances contribute. The isospin-0 amplitudes are the combinations 
of the resonance amplitudes, ${\cal M}_0^S$ and ${\cal M}_0^{f_2}$, and the non-resonating 
continuum amplitudes, ${\cal M}_0^{1C}$ and ${\cal M}_0^{2C}$~\cite{Ali:2010pq,Ali:2013xba}.   
The differential cross section for the $Y_b (10750) \to \Upsilon (nS)\, \pi^+ \pi^-$ decay 
can be found in~\cite{Ali:2010pq,Ali:2013xba}.  

\begin{figure}[tb]
\begin{center}
\includegraphics[width=0.4\textwidth]{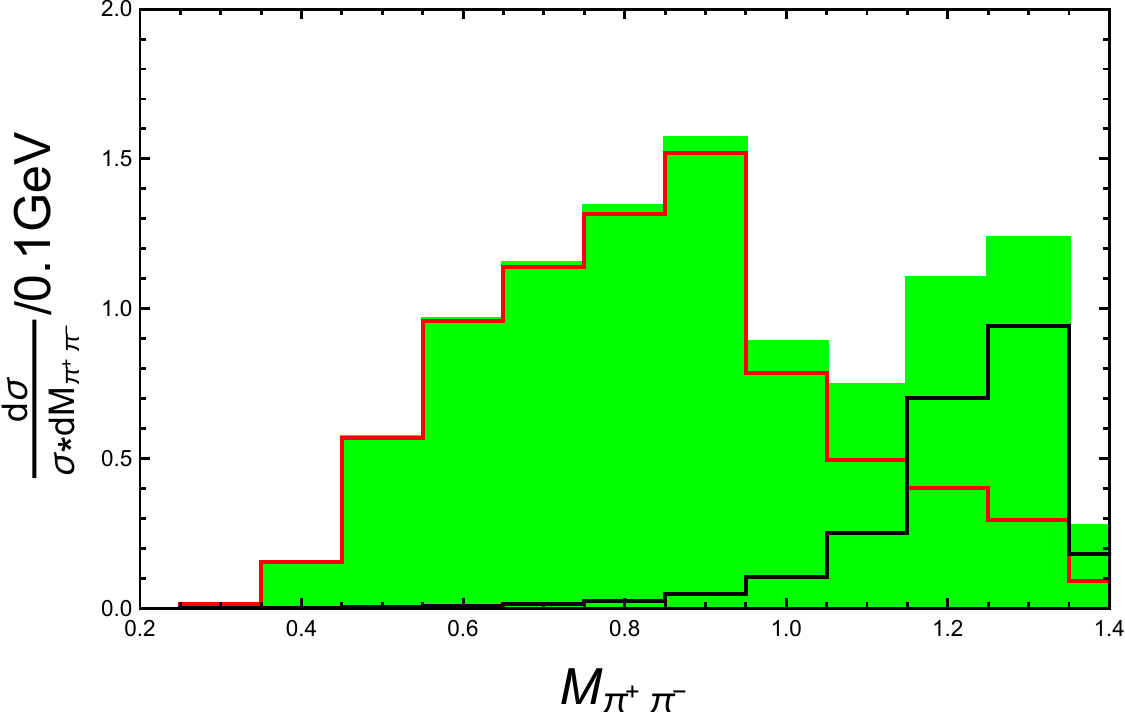}
\hfill 
\includegraphics[width=0.4\textwidth]{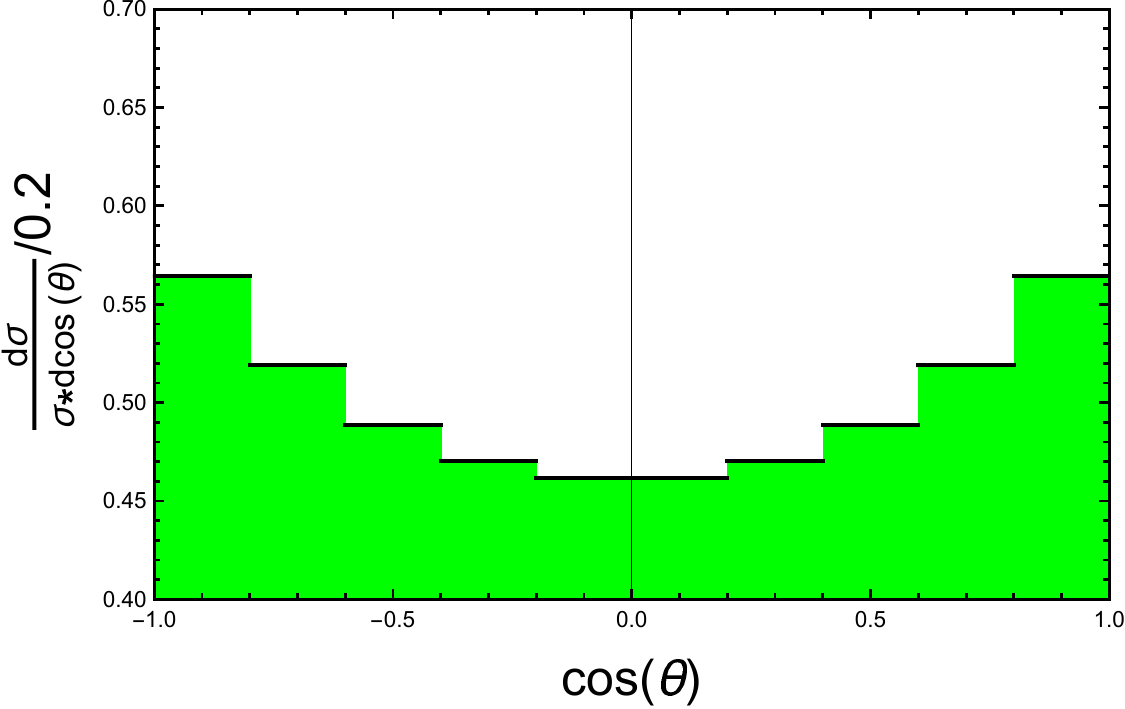}
\end{center}
\caption{\footnotesize 
The normalized resonant $m_{\pi^+ \pi^-}$~(left plot) and $\cos\theta$~(right plot) 
distributions for $e^+ e^- \to Y_b (10750) \to \Upsilon(1S) \pi^+ \pi^-$ are shown 
using the coupling constants obtained in~\cite{Ali:2010pq} (green histogram). 
The contributions from $f_0 (500)$ and $f_0 (980)$ scalars (left red curve)  
and $f_2 (1270)$ (right black curve) are indicated in the left plot.
} 
\label{fig:spectra}
\end{figure}

The $\pi^+ \pi^-$ invariant mass and angular distributions in the 
$Y_b (10750) \to \Upsilon (1S)\, \pi^+ \pi^-$ are presented in Fig.~\ref{fig:spectra}. 
They are normalized by the cross section 
$\sigma_{\Upsilon (1S)\, \pi^+\pi^-}^{\rm Belle} = (1.61 \pm 0.16)$~pb 
measured by the Belle Collaboration~\cite{Abe:2007tk}.  
Only resonant contributions are plotted, using the relevant input parameters~\cite{Ali:2010pq}.  
Spectral shapes will be modified after a realistic non-resonant contribution is included. 
A fit can only be undertaken as updated experimental measurements become available.

We also obtain estimates of the branching fractions of $Y_b (10750) \to \Upsilon (nS)\, \pi^+ \pi^-$ 
decays~\cite{Ali:2013xba}. The products $\Gamma_{ee} \times {\cal B}$ are measured by the Belle 
Collaboration~\cite{Abdesselam:2019gth},  
while $\Gamma_{ee} [Y_b (10750)] = (13.7 \pm 1.8)$~eV is known from the Belle and BaBar data analysis 
on~$R_{b \bar b}$~\cite{Dong:2020tdw}.  
Corresponding ranges of the branching fractions are as follows~\cite{Ali:2013xba}: 
${\cal B}_{\Upsilon(1S) \pi^+ \pi^-} = (0.9 - 3.4)\%$,  
${\cal B}_{\Upsilon(2S) \pi^+ \pi^-} = (3.9 - 8.9)\%$, and    
${\cal B}_{\Upsilon(3S) \pi^+ \pi^-} = (1.5 - 1.9)\%$.  
Note that due to the dominant tetraquark nature of $Y_b(10750)$, and its quark content, 
$Y_b(10750) \to B_s^{(*)} \bar B_s^{(*)}$ decays are not anticipated, in agreement with 
the data from the Belle Collaboration~\cite{Abdesselam:2016tbc}.   

\section{Summary} 

The tetraquark-based interpretation of the new structure $Y_b (10750)$ 
found in the $e^+ e^-$ annihilation by the Belle Collaboration is presented.  
The $Y_b (10750)$- and $\Upsilon (10860)$-states are assumed to be 
the tetraquark-$b \bar b$-mixed states, anticipated in the large-$N_c$ limit. 
The $b \bar b$-component is used to predict the hadroproduction and Drell-Yan cross sections at the LHC.   
A crucial test of this interpretation is in the $m_{\pi^+ \pi^-}$ and $\cos\theta$ distributions
in the $Y_b (10750) \to \Upsilon (nS)\, \pi^+ \pi^-$ decays, whose resonant contribution is worked out.  
They are not expected in other dynamical schemes such as $Y_b (10750)$ interpreted as a $D$-wave 
$b \bar b$-state, with a very large $S-D$ mixing. 
The tetraquark-$Q \bar Q$ mixing scheme suggested has wider implications.

\acknowledgments{
The work of W.\,W. is supported in part the National Natural Science Foundation of China 
under Grant Nos. 11575110, 11735010,  11911530088,  and the Natural Science Foundation of Shanghai 
under Grant No. 15DZ2272100. 
A.\,P. and W.\,W. acknowledge financial support by the Russian Foundation for Basic Research 
and National Natural Science Foundation of China according to the joint research project 
(Nos. 19-52-53041 and 1181101282).
This research is partially supported by the ``YSU Initiative Scientific Research Activity'' 
(Project No.~AAAA-A16-116070610023-3). 
}

\bibliographystyle{JHEP}
\bibliography{Contrib-Parkhomenko-ID249} 


  


\end{document}